\documentclass[aip, reprint, showkeys, superscriptaddress, numerical ]{revtex4-1}

\usepackage{amssymb,amsmath,amsfonts,amstext, geometry,marvosym, graphics,subfigure,psfrag, bm}
\usepackage[english]{babel}

\usepackage{times}
\usepackage{graphicx}
\usepackage{overpic}
\usepackage{rotating}
\usepackage{multirow}
\usepackage{hyperref}
\usepackage{accents}

\hypersetup{colorlinks = true,linkcolor = blue, anchorcolor = blue, citecolor = blue, pdfstartpage=1, filecolor = red,urlcolor = black}
\renewcommand{\figurename}{Fig.}
 
\newcommand{\reffig}[1]{Fig.~\ref{#1}}

\usepackage{color}

\def\munderbar#1{\underline{\sbox\tw@{$#1$}\dp\tw@\z@\box\tw@}}

\def\be{\begin{equation}}
\def\ee{\end{equation}}
\def\bea{\begin{eqnarray}}
\def\eea{\end{eqnarray}}
\def\r{{\bf r}}

\def\q{{\bf q}}

\def\g{{\bf \hat g}}
\def\n{{\bf \hat n}}

\def\P{{\cal P}}

\def\O{{\cal O}}

\def\lb{\left[}
\def\rb{\right]}

\def\d{\mbox{d}}

\begin{document} 

\title{Universal power-law scaling of water diffusion in human brain defines what we see with MRI }

\newcommand{\new}{\color{red}}

\makeatletter
\renewcommand\@biblabel[1]{#1.}
\makeatother

\author{Jelle Veraart}
\email[Corresponding author: ]{Jelle.Veraart@nyumc.org}
\affiliation{Bernard and Irene Schwartz Center for Biomedical Imaging, Department of Radiology, New York University School of Medicine, New York, NY, USA}
\affiliation{Center for Advanced Imaging Innovation and Research, Department of Radiology, New York University School of Medicine, New York, NY, USA}
\affiliation{iMinds Vision Lab, Department of Physics, University of Antwerp, Antwerp, Belgium}
\author{Els Fieremans}
\affiliation{Bernard and Irene Schwartz Center for Biomedical Imaging, Department of Radiology, New York University School of Medicine, New York, NY, USA}
\affiliation{Center for Advanced Imaging Innovation and Research, Department of Radiology, New York University School of Medicine, New York, NY, USA}
\author{Dmitry S. Novikov}
\affiliation{Bernard and Irene Schwartz Center for Biomedical Imaging, Department of Radiology, New York University School of Medicine, New York, NY, USA}
\affiliation{Center for Advanced Imaging Innovation and Research, Department of Radiology, New York University School of Medicine, New York, NY, USA}


\begin{abstract}
Development of successful therapies for neurological disorders depends on our ability to diagnose and monitor the progression of underlying pathologies at the cellular level. 
Physics and physiology limit the resolution of human MRI to millimeters, three orders of magnitude coarser than the cell dimensions of microns. 
A promising way to access cellular structure is provided by diffusion-weighted MRI (dMRI), a modality which exploits the sensitivity of the MRI signal to micron-level Brownian motion of water molecules strongly hindered by cell walls. 
By analyzing diffusion of water molecules in human subjects, here we demonstrate that biophysical modeling has the potential to break the intrinsic MRI resolution limits. The observation of a universal power-law scaling of the dMRI signal identifies the contribution from water specifically confined inside narrow impermeable axons, validating the overarching assumption behind models of diffusion in neuronal tissue. This scaling behavior establishes dMRI as an {\it in vivo} instrument able to quantify intra-axonal properties orders of magnitude below the nominal MRI resolution, spurring our understanding of brain anatomy and function.
\end{abstract}

\maketitle 
%
\par
\begin{figure*}
\centering 
\includegraphics[width = 0.95\textwidth]{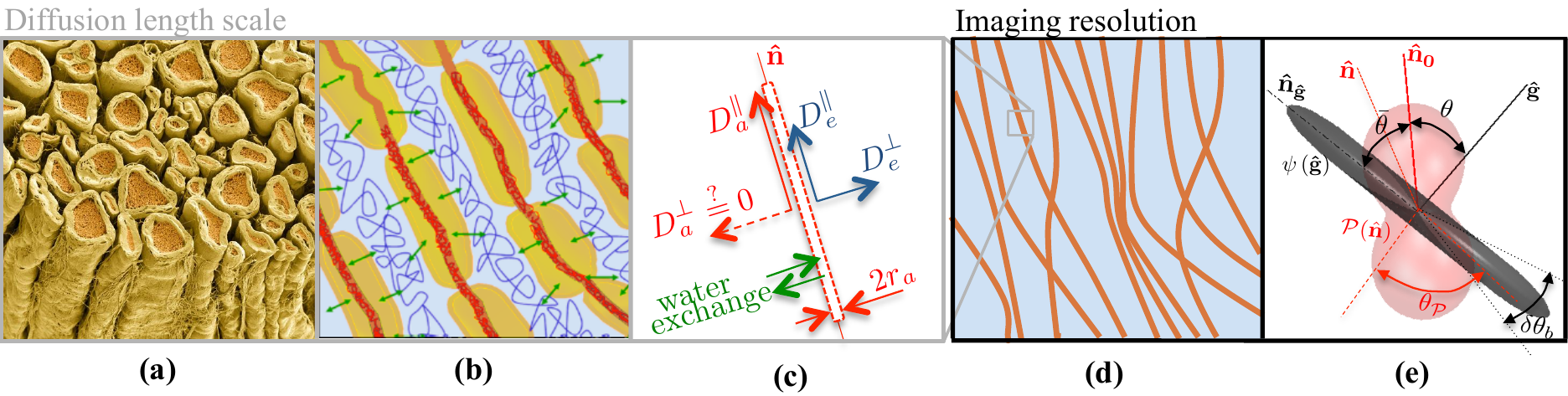}
\caption{\label{fig:model} \textbf{Microstructural origin of dMRI signal:}  The existing conjecture \cite{Behrens2003, Kroenke2004, Assaf2004, Jespersen2007, Jespersen2010, Fieremans2011, Zhang2012, lemonade-ismrm, Jensen2016} presents water diffusion in WM (a, b) as restricted diffusion in an array of axons, represented by ``sticks" (red) embedded in the hindered extra-axonal water (blue), whereas contribution of myelin water (yellow) is considered to be negligible due to its short $T_2$. (c) Schematic representation for water diffusion properties of an individual stick.  Given the diffusion time $t\sim 100\,$ms, the associated diffusion length scale of $\simeq 10\,\mu$m is at least two orders of magnitude smaller than the imaging resolution (d). Hence, the measured intra-axonal signal reflects the averaging over an ensemble of sticks, whose directions are captured by an orientational distribution function $\P(\n)$. (e) Qualitatively, only the directions within the range $\delta \theta \sim (b D_a^\parallel)^{-1/2} \ll 1$ transverse to the diffusion direction $\g$ contribute to the observed signal at large $b$, yielding the asymptotic signal scaling (\ref{funcform}) solely from intra-axonal water because the extra-axonal signal decays exponentially fast with $b$. }
\end{figure*}

The viability of model-based ``super-resolution" MRI rests on validating fundamental model assumptions.
In white matter (WM), the most essential assumption underpinning most biohysical models\cite{Behrens2003, Kroenke2004, Assaf2004, Jespersen2007, Novikov2010, Novikov2014, Burcaw2015, Fieremans2016, Jespersen2010, Fieremans2011, Zhang2012, lemonade-ismrm, Jensen2016}  is compartmentalization --- i.e. representing the dMRI signal as a sum of independent contributions from separate pools of water, corresponding to locally anisotropic intra- and extra-axonal spaces, Fig.~\ref{fig:model}. 

In particular, the defining architectural signature of neuronal tissue from the water diffusion standpoint has been the conjecture \cite{Behrens2003, Kroenke2004, Assaf2004, Jespersen2007} of the narrow impermeable channels (``sticks") representing axons (and possibly glial cell processes) inside which diffusion is locally one-dimensional. 
The proven ability to map the corresponding directional diffusion coefficients inside these ``sticks" ($D_a^\parallel$) and outside them ($D_e^\parallel$ and $D_e^\perp$), as well as the compartment water fractions, would turn dMRI into a unique non-invasive instrument able to discern between specific intra- and extra-cellular disease processes, 
such as demyelination, axonal loss, oedema and inflammation. 

However, the Achilles's heel of model-based approaches has been  
the lack of validation of underlying model assumptions.  As histology is not directly related to a diffusion measurement, it neither can quantify MRI-relevant markers of cell integrity, such as diffusion coefficients and membrane permeability, nor prove the overarching picture of ``sticks" for the neurites, Fig.~\ref{fig:model}.  

\begin{figure}
\centering 
\includegraphics[width = .43\textwidth]{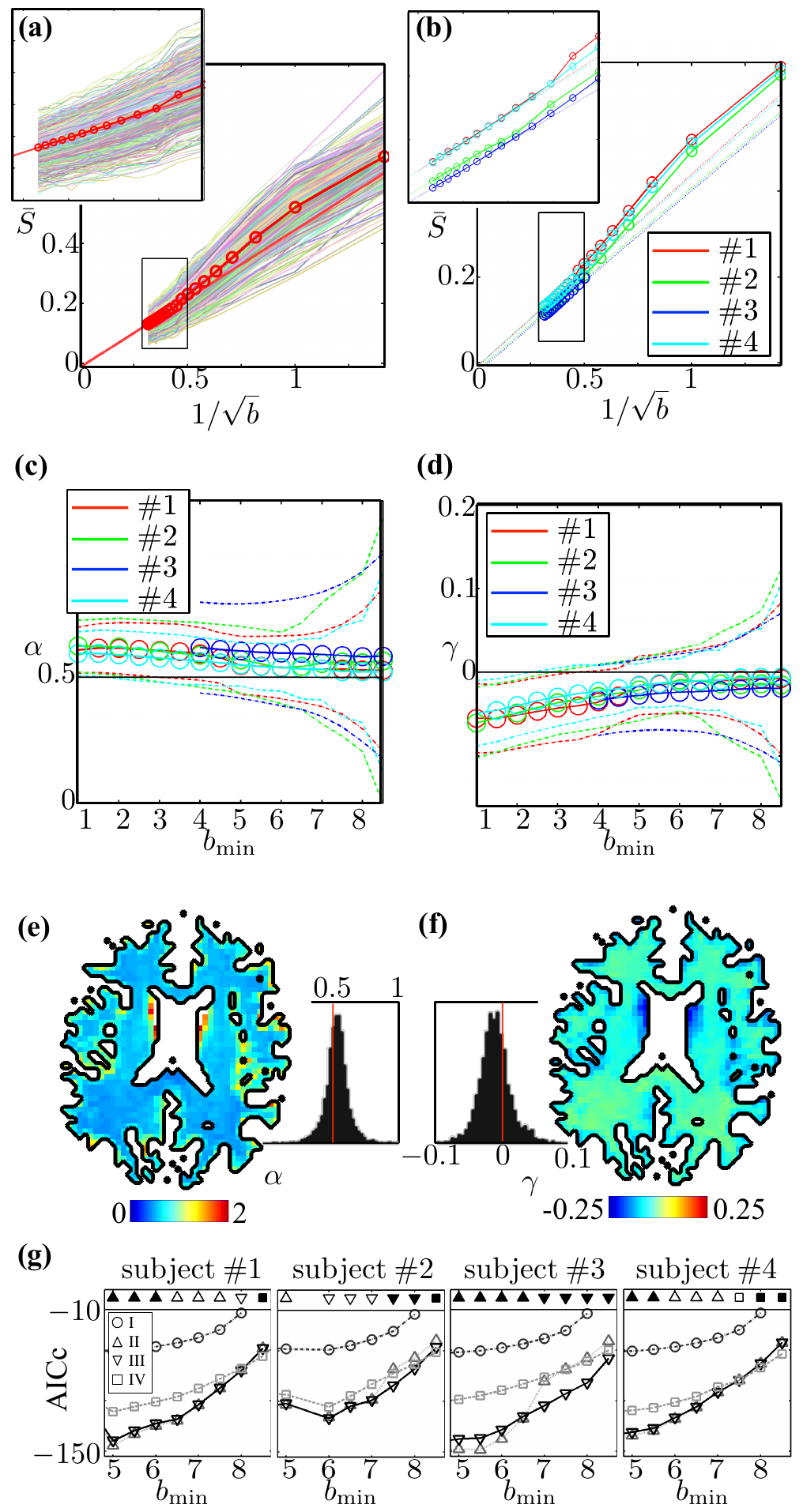}
\caption{\label{fig:scaling} \textbf{Observation of power law (1) with $\alpha=1/2$:} (a) Asymptotically linear scaling of the isotropically averaged signal as function of  $1/\sqrt{b}$ in all WM voxels for subject $\#1$. (b) For each subject, the average signal decay is shown (solid line), as well as the linear approximation for $b > 6\, \mathrm{ms/ \mu m^2}$ (dashed). (c, d)  Estimating the power exponent $\alpha$ and intercept $\gamma$ from data in the $b$-range [$b_\mathrm{min},\, 10$] for all WM voxels shows asymptotic convergence to 1/2 and 0, respectively, as function of $b_{\rm min}$. The mean and 95\% confidence intervals are plotted in solid and dashed lines. (e, f) The maps and histograms of $\alpha$ and $\gamma$ for subject $\#1$ and $b_\mathrm{min} = 7\, \mathrm{ms/ \mu m^2}$ are shown. (g) Model selection via the AICc indicates that models with fixed $\alpha=1/2$ should be preferred over the others in case of $b_\mathrm{min} > 6 \,\mathrm{ms/ \mu m^2}$. 
}
\end{figure}

Here we argue that our experimental {\it in vivo} observation of the universal power-law form (Fig.~\ref{fig:scaling})
\begin{equation}
\label{funcform}
{S}(b \rightarrow \infty) \simeq \beta \cdot b^{-\alpha} + \gamma 
\end{equation}
of the dMRI signal $S$, with exponent $\alpha = 1/2$, in the human brain validates for the first time the key miscrostructural assumptions behind models of diffusion in  WM: That one-dimensional ``sticks" (axons) form a universal anisotropic diffusion compartment; the exchange between intra- and extra-axonal water is not relevant; and that the fraction $\gamma$ of fully restricted water is negligible in the clinically accessible regime. 

The dMRI signal $S = \int\! \d\r\, e^{-i\q\r}G_{t,\r}$ is the Fourier transform of the diffusion propagator $G_{t,\r}$ averaged over all water molecules within an imaging voxel \cite{Novikov2010},  
and the diffusion weighting\cite{Basser1994} parameter $b=q^2 t$. Here we use a diffusion time $t\approx 50\,$ms as available on clinical scanners, and vary $q$, achieving an order of magnitude greater weightings than typical $b\sim 1\,\mathrm{ms/\mu m^2}$ used in the clinic. 

%
The asymptotic power-law  (\ref{funcform}) with exponent $\alpha=1/2$ can only originate from the {\it intra-axonal water}. Indeed, consider the dMRI signal (henceforth normalized to $S|_{b=0}\equiv 1$)
\be \label{S=Pdn}
S(\g, b) = f\!\! \int\! \d\n\, \P(\n)\,  \psi_\n(\g, b) \, + \, \gamma \, + \, S^{\rm eas}(\g, b)  \quad
\ee
in the unit direction $\g$. 
The first term comes from the collection of sticks representing axons and possibly glial cell processes, with net water fraction $f$, and parameterized by the orientational distribution function (ODF) $\P(\n)$, Fig.~\ref{fig:model}. 
If the overarching brain dMRI modeling assumption \cite{Behrens2003, Kroenke2004, Assaf2004, Jespersen2007, Jespersen2010, Fieremans2011, Zhang2012, lemonade-ismrm, Jensen2016} is correct, the  signal (``stick response function'') 
$\psi_{\n}\left(\g,b\right) $ from water confined within a stick pointing in the direction $\n$, can be approximated by a simple Gaussian one-dimensional diffusion propagator $\psi_{\n}\left(\g,b\right) \equiv e^{-bD_a^\parallel (\g\n)^2}$.  
In the limit $bD_a^\parallel \gg 1$, this response function yields a non-negligible contribution only from axons falling within a thin ``pancake" $|\n\cdot\g|\lesssim (bD_a^\parallel)^{-1/2}$ {\it nearly transverse to} $\g$, Fig.~\ref{fig:model}(e), whose thickness scaling as $b^{-1/2}$ results in the  asymptotic form (\ref{funcform}).

It is essential, for the power-law scaling (\ref{funcform}) to hold and to originate solely from ``intra-stick" water,  that the dMRI signal exactly transverse to a stick, $\g\perp\n$, is not suppressed: $\psi_{\n\perp\g}$ does not decay at large $b$, equivalent to a negligible transverse diffusion coefficient $D_a^\perp$ and axonal radius (Fig.~\ref{fig:model}). 
In contrast, the extra-axonal contribution $S^{\rm eas}(\g, b) \sim e^{-bD_e(\g)}$, 
coming from water diffusion in a simply-connected space characterized by a finite diffusion coefficient $D_e(\g)$ in any direction $\g$ (with $D_e(\g) \geq D_e^\perp$, Fig.~\ref{fig:model}), decays exponentially faster than the intra-axonal signal, by virtue of $e^{-bD_e^\perp} \ll 1$ for large $b$, and can be eventually neglected, as further shown below. 
Finally, $\gamma \equiv S|_{b=\infty}$ is the possible contribution of immobile (fully restricted) water, which we will show to fall  below our detection threshold.  It is clear from the above argument that either a finite axonal radius, or a notable exchange rate between intra- and extra-axonal water, would destroy the very particular $b^{-1/2}$ scaling (\ref{funcform}). 

Figure \ref{fig:scaling}(a) demonstrates the asymptotic behavior (\ref{funcform}) based on diffusion measurements in all WM voxels (colored lines) with $0\leq b \leq 10\, \mathrm{ms/\mu m^2}$ (see {\it Methods} section for details of signal processing, minimizing signal biases due to scan drift, imaging artifacts, and correcting for the Rician noise floor \cite{Gudbjartsson1995}). 
Here, the signal was averaged over 64 diffusion directions $\g$ evenly distributed on a sphere for each $b$ to boost the signal-to-noise ratio, with an added advantage being the cancellation of the ODF shape\cite{Kaden2015}, 
$\overline{S}(b) = \int\!  \d\n \,\P(\n) \, \int\! \d\g \, e^{-bD_a^\parallel (\g\n)^2} 
\simeq {\sqrt{\pi} \over 2} (b D_a^\parallel)^{-1/2}$, since $\int\! \d\g \, e^{-bD_a^\parallel (\g\n)^2}$ is independent of fiber direction $\n$, and the ODF is normalized to $\int\! \d\n \, \P(\n) \equiv 1$. This yields the prefactor 
$\beta = \sqrt{\pi/4} \cdot f/\sqrt{D_a^\parallel}$ in equation (\ref{funcform}) for the direction-averaged signal in terms of the intra-axonal parameters $f$ and $D_a^\parallel$. 
As an average of power-law contributions (over directions and voxels) with the same $\alpha$ yields the same power law, the asymptotic behavior (\ref{funcform}) becomes most pronounced for the WM- and direction-averaged signals (red line in Fig. \ref{fig:scaling}(a)), and is reproducible in every subject, Fig. \ref{fig:scaling}(b).

Fit robustness with respect to the range of $b$, Fig.~\ref{fig:scaling}(c-f), and to the number of degrees of freedom, Fig.~\ref{fig:scaling}(g), was evaluated by considering the full and nested models to equation (\ref{funcform}) for the ranges $b \geq b_\textrm{min}$, with $b_\textrm{min}$ varying between 0 and $8.5\, \mathrm{ms/\mu m^2}$. More specifically, the evaluated models were (I) $\beta b^{-\alpha} + \gamma$ Eq.~(\ref{funcform}), (II) $\beta b^{-\alpha}$, (III) $\beta b^{-1/2} + \gamma$, and (IV) $\beta b^{-1/2}$.  We compared the relative fit quality of the nested models  by means of the corrected Akaike information criterion \cite{BurnhamAndersonBook}, see {\it Methods}.  Fig.~\ref{fig:scaling}(g) shows that fixing $\alpha = 1/2$, models (III) or (IV), leads to better fit quality than constraining $\gamma$, as long as $b \gtrsim 6\, \mathrm{\mu m^2/ms}$. 

Intersubject variability and $b_\mathrm{min}$-dependence is shown in Fig.~\ref{fig:scaling}(c,d). The  exponent $\alpha_\mathrm{II}$ and intercept $\gamma_\mathrm{III}$ have both been evaluated as function of $b_\textrm{min}$, for all WM voxels, and for all subjects. Overall, $\alpha \to 1/2$ and $\gamma \to 0$. 

Histograms and maps of the $\alpha_\mathrm{II}$ and $\gamma_\mathrm{III}$ for subject $\#1$ and $b_\mathrm{min} = 7\, \mathrm{ms / \mu m^2}$  are shown in \reffig{fig:scaling}(e-f). The maps look fairly homogenous in the WM with $\alpha_\mathrm{II}$ and $\gamma_\mathrm{III}$ values centered around 1/2 and 0, respectively, which suggests that the observed power law behavior (\ref{funcform}) with 
exponent $\alpha=1/2$ and negligible immobile water fraction $\gamma$ is {\it universal} in the human WM {\it in vivo}. This observation validates for the first time  the overarching biophysical picture  that has been the cornerstone conjecture for numerous diffusion MRI models \cite{Behrens2003, Kroenke2004, Assaf2004, Jespersen2007, Jespersen2010, Fieremans2011, Zhang2012, lemonade-ismrm, Jensen2016} since 2003: Axons (and more generally, neurites) are represented by one-dimensional ``sticks", with negligible transverse intra-axonal diffusivity; negligible exchange between intra- and extra-axonal water; and negligible immobile water contribution.

\begin{figure}
\centering 
\includegraphics[width = .5\textwidth]{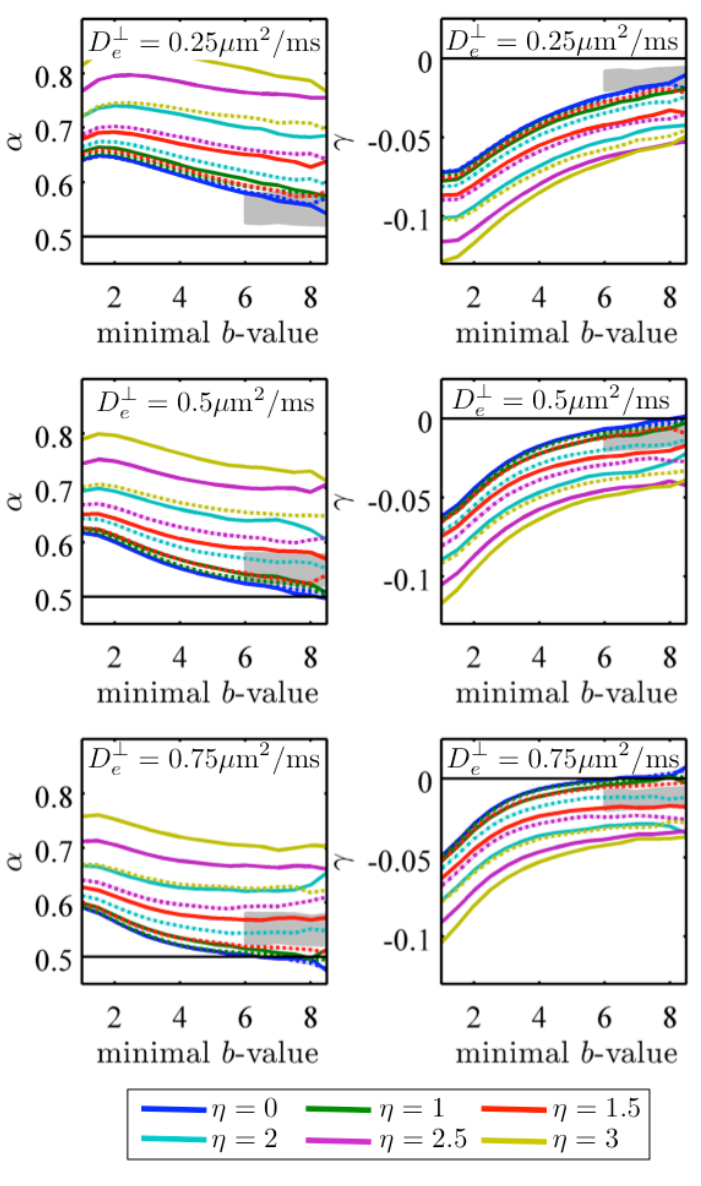}
\caption{\label{fig:sim} {\bf Effects of the residual extra-axonal diffusion and finite axonal radii.} Simulations with varying $D_e^\perp$ demonstrate the feasibility to identify the power law exponent $\alpha = 1/2$ if  axons have radii according to values reported in the histology studies of Aboitiz et al. \cite{Aboitiz1992} (solid lines) or Caminiti et al. \cite{Caminiti2009} (dashed lines), when corrected by a shrinkage factor between $\eta$ = 1 and 2 (cf. Methods).
Comparison with Fig.~\ref{fig:scaling}(c,d) (observed curves falling within the greyed-out area) excludes strong tissue shrinkage, $\eta\gtrsim 2$. 
}
\end{figure}

In Fig.~\ref{fig:sim} we argue that a slight, yet significant (significance level of 0.05) bias in the estimation in $\alpha$ and $\gamma$  is expected due to residual exponentially decaying extra-axonal signal, i.e. a signature of strongly hindered radial extra-axonal diffusion ($D_e^\perp \approx 0.5\,\mathrm{\mu m^2/ms}$) \cite{Burcaw2015,Fieremans2016},  and that the biases cannot be attributed to finite axonal radii (cf. {\it Methods} section for details of simulations).  Consequently, we conclude that our measurement is practically insensitive to axonal radii as long as they fall within the range obtained by histology \cite{Aboitiz1992,Caminiti2009}.  

The insensitivity to axonal radii should help resolve the on-going debate\cite{Horowitz2015, Innocenti2015, Burcaw2015} about the feasibility of {\it in vivo} axonal diameter mapping in the brain.   Histological studies extensively reported axonal diameters $2r$ to be in the range $0.5 - 2\,\mu$m for human WM \cite{Aboitiz1992, Caminiti2009},  with only 1\% of all axons having a diameter larger than 3$\,\mu$m \cite{Caminiti2009}, while MRI-derived axonal diameters fall in the range $3.5 -15\,\mu$m \cite{Alexander2010, Horowitz2015, Huang2015}. 
On the MRI side, the bias has been attributed to the volume-weighted contributions amplifying the tail of the distribution \cite{Alexander2010,Burcaw2015}, to the wide diffusion pulses reducing the effect of signal attenuation\cite{Vangelderen1994, Burcaw2015}, and to the effect of residual time-dependence of extra-axonal diffusion $D_e^\perp(t)$ overshadowing the relatively small $D_a^\perp \simeq r^2/4t$ \cite{Burcaw2015,Fieremans2016}. On the other hand, shrinkage during tissue fixation has been suggested as a potential shortcoming of histology\cite{Horowitz2015}, implying that {\it in vivo} axons are thicker than their histologically reported values.   
Fig.~\ref{fig:sim} shows that the effect of strong tissue shrinkage by a factor $\eta\gtrsim 2$, for which the large MRI-derived axonal radii\cite{Alexander2010,Horowitz2015} could make sense, leads to a qualitatively different form of $S(b)$ with $\alpha$ notably exceeding $1/2$, and  unphysical $\gamma<0$, both incompatible with our measurement. 


\begin{figure*}
\centering 
\includegraphics[width = \textwidth]{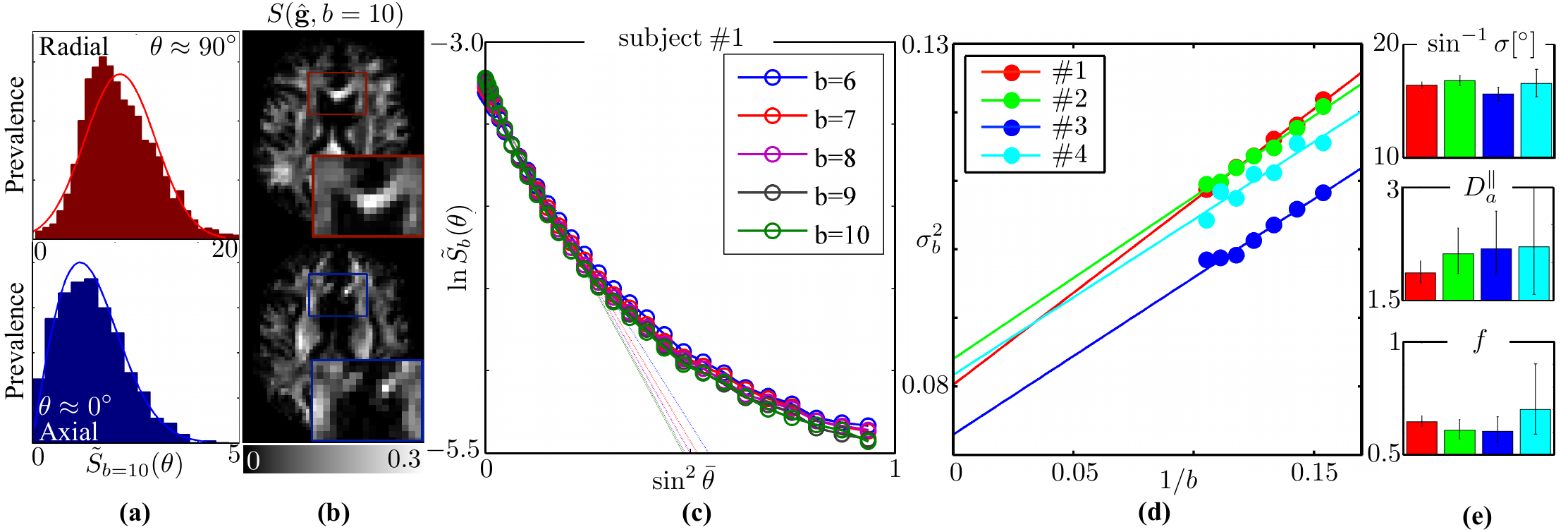}
\caption{\label{fig:dispersion} \textbf{Directional dependence of the high-$b$ signal:} (a,b) Full suppression of $\tilde{S}_b(\theta)$ is observed in the directions approximately parallel to the underlying axonal fibre (axial); signal statistics for  $\theta \leq 20^\circ$ (blue bars) obeys the Rayleigh distribution, i.e. SNR=0 (solid line). In the radial directions (red), SNR $\gtrsim 5$ even at $b=10\,\mathrm{ms/\mu m^2}$. The distribution of the probability-normalized distribution of $\tilde{S}_b(\theta)$ as function of $\sin^2 \bar\theta$ is non-Gaussian (c) and its variance $\sigma_b^2$ (computed as the slope of $\ln \tilde{S}$ vs $\sin^2 \bar\theta$, panel (c)) decays as $1 / b$ (d). The intercept and slope of the variance as function of $1 / b$ returns an {\it in vivo} estimate of the intrinsic axonal orientational dispersion $\sigma^2$ and intra-axonal diffusivity $D_a^\parallel$, respectively, panel (e).  Axonal water fraction $f$ is estimated using $D_a^\parallel$ and the parameter $\beta$ from equation (\ref{funcform}). Error bars span the $95\%$ confidence interval; bar plots show reproducibility over 4 subjects. 
} 
\end{figure*}

Figure \ref{fig:dispersion} shows the strong directional dependence of the signal $S(\g, b)$, providing additional evidence for the lack of an isotropically restricted component, i.e. $\gamma\to 0$. Here, we select voxels characterized by a single fiber population (SFP) \cite{Tax2014}, and focus on the signal $\tilde S_b(\theta)$ as a function of the angle $\theta$ between the gradient direction $\g$ and the principal fiber direction $\bf \hat n_0$ (determined as a principal diffusion tensor direction).  
To increase precision, we averaged $\tilde S_b(\theta)$ over all SFP voxels, normalizing voxel-wise contributions by the spatially-dependent noise level $\sigma(x)$ \cite{Veraart2016}. We observe a remarkably high signal-to-noise ratio $\mathrm{SNR \gtrsim 5}$ even for an ``extreme" diffusion weighting $b=10\,\mathrm{ms/\mu m^2}$ in the radial direction ($\theta \approx 90^\circ$, $\g \perp \n_0$). We emphasize that it is the presence of effectively zero-radius ``sticks" that explains any visible anatomy in Fig.~\ref{fig:dispersion}b; in non-neuronal tissues, with large and non-``stick-like" cells, at $b=10\,\mathrm{ms/\mu m^2}$ and $D\sim 1\,\mathrm{\mu m^2/ms}$, one would observe pure noise since for such parameters, $e^{-bD}\sim 10^{-4}$.

Conversely, the signal $\tilde S_b(\theta)$ in the axial direction, binned within the cone $\theta \leq 20^\circ$ (when $\g$ is almost parallel to the principal fiber direction), is {\it fully suppressed}, as it reaches the Rician noise floor $\tilde S = \sqrt{\pi}/2$ for the magnitude MR images. Furthermore, the axial signal statistics is precisely governed by the Rayleigh distribution (Rice distribution with zero signal) \cite{Veraart2016}, the blue line in Fig.~\ref{fig:dispersion}(a) drawn without any adjustable parameters, corroborating the accuracy of our noise estimation method and our conclusion about the unobservable $\gamma$.  

The presence of isotropic immobile water has been conjectured \cite{Stanisz1997} in 1997, as water possibly trapped inside the small glial cells such as the oligodendrocytes, and other small  compartments (e.g. vesicles). However, the negligible $\gamma$ and the pure-noise statistics in the axial direction indicates that \textit{in vivo} human diffusion MRI is practically insensitive to such contributions, either because their volume fraction is too small, or because their $T_2$ relaxation time is too short, or because the water exchange rate is too fast on the scale of our diffusion time $t \approx 50\,\mathrm{ms}$ for treating them as coming from separate compartments.

Finally, we use the directional signal dependence $\tilde S_b(\theta)$ in order to determine an SFP-averaged fiber ODF $\tilde \P(\n)$, Fig.~\ref{fig:dispersion}c, and to obtain estimates of the intra-axonal diffusivity $D_a^\parallel$ and axonal orientational dispersion (i.e. the degree of their misalignment within SFP voxels). Our method rests on the following intuition: For $b D_a^\parallel \to \infty$, the stick response function becomes infinitely sharp, such that the corresponding $\tilde S_{b\to \infty}(\theta) \propto \tilde \P(\bar \theta)$, where the complementary angle $\bar \theta = \pi/2 - \theta$, Fig.~\ref{fig:model}e. Since we bin the distribution $\tilde S_b(\theta)$ only as function of the polar angle $\theta$ to reduce the effects of noise, our average SFP ODF $\tilde \P(\n) \equiv \tilde \P(\bar\theta)$ will be axially symmetric. 
Fig.~\ref{fig:dispersion}c shows the gradual sharpening of $\tilde S_b(\theta)$, illustrating the approach toward the intrinsic $\tilde \P(\bar\theta)$, for the increasing $b$. 
By focusing on the small-$\bar \theta$, Gaussian part $\tilde P(\bar \theta) \sim e^{-\sin^2\bar\theta/2\sigma^2}$ of the ODF, we can deconvolve its intrinsic variance $\sigma^2$ from a sequence of finite-$b$ measurements. In {\it Methods}, we prove that in this approximation, 
the signal $\tilde S_b(\theta) \sim e^{-\sin^2\bar\theta/2\sigma_b^2}$ approaches the Gaussian shape with the variance $\sigma_b^2 = \sigma^2 + \frac1{2bD_a^\parallel}$, i.e. the intrinsic variance and that from the stick response function add up, as it could be intuitively expected. 
In Fig.~\ref{fig:dispersion}d, we observe that this variance $\sigma_b^2$, calculated via the slope of $\tilde S_b$ with respect to $\sin^2\bar\theta$ around $\bar\theta=0$, indeed scales linearly with $1/b$ in all subjects. 
This scaling of the variance is consistent with the signal scaling (\ref{funcform}), confirming our picture of asymptotic sensitivity only to the intra-axonal water within narrow impermeable sticks, and further allows us to determine both the intrinsic fiber orientational dispersion $\sigma$, and the intra-axonal diffusivity $D_a^\parallel$ separately from the axonal water fraction $f$. This shows the added value of studying the directional signal variance, as compared to the overall signal magnitude for a given direction, or the directional average $\overline{S}$ (Fig.~\ref{fig:scaling}), in which these quantities are mixed in the parameter $\beta \propto f/\sqrt{D_a^\parallel}$.  We estimate the intra-axonal diffusivity $D_a^\parallel$ to be in the range [1.9, 2.2] $\mathrm{\mu m^2 /ms}$, whereas the axonal water fraction $f$ ranges between 0.6 and 0.7 amongst the four subjects (Fig.~\ref{fig:dispersion}e).  
Extrapolating the observed linear function to the $1/b=0$ intercept provides an \textit{in vivo} estimate of intrinsic ODF dispersion $\sigma$. The estimated dispersion  
angle $\sin^{-1}\sigma \approx 17^\circ$ in all subjects is in excellent agreement with previous histological studies yielding dispersion of about $18^\circ$  \cite{ronen2014, leergaard2010}.

To conclude, this study represents the first observation of a scale-invariant functional form $S\sim b^{-1/2}$ for an \textit{in vivo} human diffusion-weighted MRI signal. 
Experimental detection of specific functional forms -- a validation paradigm borrowed from the physical sciences -- is instrumental for selecting\cite{Novikov2014,Burcaw2015,Fieremans2016} the family of plausible microstructural models, and for non-invasively validating their assumptions.  
The remarkably slow decay (\ref{funcform}) of the signal, retaining much SNR even for very high $b$, provides an exciting avenue for probing brain tissue microstructure with extremely strong diffusion gradients on clinical systems, such as on {\it Human Connectom} scanners, potentially enabling much higher resolution in fiber tractography used in mapping brain anatomical connectivity and in the presurgical planning, as well as higher precision in estimating biophysical parameters\cite{Behrens2003, Kroenke2004, Assaf2004, Jespersen2007, Jespersen2010, Fieremans2011, Zhang2012, lemonade-ismrm, Jensen2016} of microstructural tissue integrity, thereby fostering the translation of advanced diffusion MRI methods into basic neuroscience research and clinical practice. 


\section*{Methods}

\quad \quad \textbf{MRI:} Four healthy volunteers underwent imaging on a Siemens Prisma 3T MR scanner, equipped with a $80 \,\mathrm{mT/m}$ gradient system, after obtaining informed consent, using a 64-channel receiver head coil. The body coil was used for transmission. An echoplanar readout diffusion-weighted sequence was used to acquire the dMRI data. Diffusion weighting was applied along 64 isotropically distributed gradient directions for each of the 21 $b$-values that were equidistantly distributed in the range $[0, 10\, \mathrm{ms/\mu m^2}$]. Following imaging parameters were kept constant throughout the data acquisition sequence: $\mathrm{TR/TE}: 4000/105\,\mathrm{ms}$,  matrix: $80 \times 80$, NEX: 1, in-plane resolution:  $3 \times 3 \,\mathrm{mm^2}$, slice thickness: $3\,\mathrm{mm}$, slices: 38, parallel imaging: GRAPPA with acceleration factor 2, reconstructed using the adaptive combine  algorithm to ensure Rician data distribution, multiband acceleration with factor 2, and no partial Fourier. 

\textbf{Image processing: } MPPCA noise estimation and denoising\cite{Veraart2016} allowed to strongly reduce the noise in the data and to estimate the noise map $\sigma(x)$ by exploiting the inherent redundancy in diffusion MRI data. The positive signal bias, inherent to low-SNR magnitude MR data, was removed by using the method of moments \cite{Koay2006b}, where the denoised signal was used as a proxy for the Rician expectation value.
Denoised and Rice-floor-corrected images were subsequently corrected for Gibbs ringing \cite{Kellner2015, Veraart2015}, geometric eddy current distortions and subject motion \cite{Smith2004, Glasser2013}.
Using FSL's FAST algorithm \cite{Smith2004}, an initial WM mask was extracted from the $b=0$ images.  To avoid voxels affected by partial voluming with the grey matter, a more conservative segmentation was obtained by omitting all voxels with a fractional anisotropy smaller than 0.6. 

\textbf{Model selection: } The corrected Akaike's information criterion\cite{BurnhamAndersonBook} (AICc) was used to compare the relative fit quality of following nested models: (I) $\beta b^{-\alpha} + \gamma$ [equation (\ref{funcform})], (II) $\beta b^{-\alpha}$, (III) $\beta b^{-1/2} + \gamma$, and (IV) $\beta b^{-1/2}$. Model (II) was linearized by the log transformation, i.e. $\ln S(b \geq b_\textrm{min}) = \ln \beta - \alpha \ln b$, and fitted using linear regression. The models' degrees of freedom were 3, 2, 2, and 1, respectively. Fig.~\ref{fig:scaling}(g) shows that the decreased complexity of models (II), (III), or (IV) should be preferred over the full model (I). The row of markers on top of the graphs indicates which model has the lowest AICc for a given minimal $b$-value. Significant differences, i.e. $\Delta \mathrm{AICc} > 2$, are indicated by black markers.  AICc differences between the nested models (II), (III), and (IV) are typically small, and even not always statistically significant, i.e. $\Delta \mathrm{AICc}<2$. However, in case of significant differences, fixing $\alpha = 1/2$, i.e. for models (III) or (IV), always leads to better fit quality than constraining $\gamma$, as long as $b \gtrsim 6\, \mathrm{\mu m^2/ms}$. 

\textbf{Noise propagation and effect of finite axonal radii:} In order to evaluate the feasibility to detect the $\alpha = 1/2$ power law  scaling,  Rice-distributed synthetic data  was generated for the same imaging protocol using a two-compartmental  signal-generating model of orientationally dispersed WM ($\gamma=0$).  Without loss of generality, we choose $\P(\n)$  to be the axially symmetric Watson distribution:  $\P(\n) = M\left(1/2, 3/2, \kappa \right)^{-1} e^{\kappa (\g\n)^2}$,
where $M$ is a confluent hypergeometric function and $\kappa$ is the concentration parameter that describes the axonal dispersion.  The hindered extra-axonal signal
\begin{equation}
S^{\rm eas}(\g, b) = (1-f) e^{-bD_e^\perp} \int\! \d\n\, \P(\n) e^{-b(D_e^\parallel-D_e^\perp)(\g\n)^2}
\end{equation}
is parameterized by axial diffusivity $D_{e}^\parallel$ and transverse diffusivity $D_{e}^\perp$, relative to each axonal fiber in direction $\n$, Fig.~\ref{fig:model}.  $D_{e}^\parallel$ was sampled from a normal distribution with mean$\pm$sd $2\pm0.2\,\mathrm{\mu m^2/ms}$, whereas the effect of varying $D_{e}^\perp$ was evaluated by sampling from distributions with mean values $\bar{D}_{e}^\perp$ = 0.25, 0.50, 0.75, and sd = 0.1. The  intra-axonal compartment had $f = 0.65 \pm 0.10$,  $D_a^\parallel = 2\pm0.2\,\mathrm{\mu m^2/ms}$ (cf. Fig.~\ref{fig:dispersion}e), Watson concentration parameter $\kappa$ ranging from 0.1 to 30, and  finite axon radii $r_i$. Radial signal attenuation within the impermeable cylinders\cite{Vangelderen1994} accounted for the axonal radii distribution by adding the signals for every $r_i$ weighted by $ r_i^2$, using $r_i$ from the bins of the measured distributions\cite{Aboitiz1992,Caminiti2009}.  
To account for possible tissue shrinkage in histology, we multiplied the digitized histograms\cite{Aboitiz1992,Caminiti2009} with a uniform shrinkage factor $\eta = \{1, 1.5, 2, 2.5, 3\}$. We also used $\eta =0$ to simulate the effect of zero-radius axons.  
Note that the shrinkage factor is histologically accepted to be at most $\approx 30\%$ ($\eta \approx 1.43$). No immobile water was added. For each $\eta$, 1000 ``voxels" were simulated by sampling the ground truth values in their respective intervals, with ${\rm SNR} = 30$ for $S(b=0)$, to mimic our \textit{in vivo} data sets.  Processing of the simulated data included denoising and Rician bias correction, as described above. 

\textbf{Stick response deconvolution:} Assuming fiber tract direction $\n_0 \parallel {\bf \hat z}$ and a sufficiently narrow  axially-symmetric 
ODF $\P(\n)$, we can approximate it by the Gaussian (Watson) shape 
in the vicinity of ${\bf \hat z}$. In this limit, the intra-axonal contribution to the signal 
\begin{align}
S(\g,b) &= f c_\kappa \int\!\d\n \, e^{-\kappa(n_x^2 + n_y^2) - bD_a^\parallel(\g\n)^2}  \nonumber \\
&= 
2fc_\kappa \int_{-\infty}^\infty\! {\d\lambda\over 2\pi}\, e^{i\lambda} \int\! {\d^3{\bf n} \over 4\pi}\, e^{-{\bf n} {\cal A} {\bf n}} \nonumber \\
&=
{fc_\kappa \over 2\pi} \int_{-\infty}^\infty\! {\d\lambda\over 2\pi}\, {\pi^{3/2} \, e^{i\lambda} \over \det^{1/2}{\cal A}(\lambda)} \nonumber
\end{align}
where the ODF normalization $c_\kappa \simeq 2\kappa$ 
follows from 
$c_\kappa^{-1} = \int_0^{\pi/2}\! \sin\theta \d\theta \, e^{-\kappa\sin^2\theta} \simeq \int_0^\infty\! \theta\d\theta\, e^{-\kappa\theta^2} = \frac{1}{2\kappa}$
 in the narrow-ODF limit $\kappa \gg 1$, and
in the second equality we introduced the integration over the 3d space of ${\bf n}$ using the constraint 
$\delta(|{\bf n}|-1) = 2 \delta({\bf n}^2 - 1) \equiv 2\int\! {\mathrm{d}\lambda\over2\pi}\, e^{i\lambda(1-{\bf n}^2)}$, 
such that the symmetric matrix 
${\cal A} = i\lambda + bD_a^\parallel \,\g\otimes\g  + \kappa ( 1 - {\bf \hat z} \otimes {\bf \hat z})$. 
The 3d Gaussian integration is performed exactly by means of a unit-Jacobian orthogonal transformation ${\bf n} \to {\cal R}{\bf n}$ diagonalizing ${\cal A}$, yielding the last equality. 
Using 
\begin{align}
\det {\cal A} &= (i\lambda+\kappa)\lb (i\lambda)^2 + (bD_a^\parallel+\kappa)\cdot i\lambda + \kappa bD_a^\parallel g_z^2\rb \nonumber \\
& \equiv (i\lambda+\kappa)(i\lambda + x_-)(i\lambda+x_+) \nonumber
\end{align}
where $2x_\pm = bD_a^\parallel+\kappa \pm \sqrt{(bD_a^\parallel+\kappa)^2 - 4\kappa bD_a^\parallel g_z^2}$, and deforming the integration contour from the real axis into the upper 
half-plane of the complex variable $\lambda$ according to the Jordan's lemma, the resulting integration encircles the two branch cuts: between 
$ix_- \simeq i\kappa bD_a^\parallel g_z^2/(\kappa+bD_a^\parallel)$ and $i\kappa$, and between $ix_+ \simeq i(bD_a^\parallel+\kappa - x_-)$ and $i\infty$, along the positive imaginary axis. Parametrizing $\lambda = iy$, we obtain

\begin{align} \nonumber
S(\g,b) = &{fc_\kappa \over 2\sqrt{\pi}} 
\lb 
\int_{x_-}^\kappa {\d y \, e^{-y} \over \sqrt{(\kappa-y)(y-x_-)(x_+-y)}} \right. \\
&\left. -\int_{x_+}^\infty {\d y \, e^{-y} \over \sqrt{(y-\kappa)(y-x_-)(y-x_+)}}   \nonumber
\rb .
\end{align}
Our goal is to find the dependence of the signal on $g_z = \cos\theta \equiv \sin \bar \theta$ (Fig.~\ref{fig:model}) in the limit $bD_a^\parallel \gtrsim \kappa \gg 1$, i.e. when our stick response function is sharper than the ODF, and the ODF is sufficiently sharp to justify using the Gaussian (Watson) shape around its apex. 
In this limit, the second term is exponentially suppressed as $\sim e^{-bD_a^\parallel}$ and can be neglected in comparison with the first term.
The first term  
is dominated by the region around its lower bound $x_- \simeq \kappa g_z^2$ extending up to $y\lesssim 1$ 
due to the exponentially decaying weight $e^{-y}$. Extending the upper bound from $y=\kappa$ to $y=\infty$ would result in an exponentially negligible error; this in turn makes the resulting integral exactly solvable after neglecting the $y$-dependence under the square root, except around $y=x_-$:   
\begin{align} \nonumber
S(\g,b) & \simeq {fc_\kappa \over 2\sqrt{\pi \kappa (x_+  - x_-)}} \int_{x_-}^\infty {\d y\, e^{-y} \over \sqrt{y-x_-}}  \\
&= {c_\kappa \over 2\sqrt{\kappa}} {e^{-x_-}\over \sqrt{x_+ - x_-}}.  \nonumber
\end{align}
As a result, we obtain the asymptotically Gaussian shape of the direction-dependent signal
\be \label{deconv}
S(\g,b) \equiv  S_b(\bar\theta) \simeq {f \over \sqrt{1 + bD_a^\parallel/\kappa}}\ e^{-\sin^2\bar\theta/2\sigma_b^2} 
\ee
with $\sigma_b^2 = \frac1{2\kappa} + \frac1{2b D_a^\parallel}$, where the intrinsic ODF variance $\sigma^2 = \frac1{2\kappa}$ is increased by that from the stick response function width, scaling as 
$\frac1{2b D_a^\parallel} \sim \frac1b$. 

\bibliography{refs}

\section*{Acknowledgement}
JV is a Postdoctoral Fellow of the Research Foundation - Flanders (FWO; grant number 12S1615N). 
EF and DSN were supported by the Fellowship from Raymond and Beverly Sackler Laboratories for
Convergence of Physical, Engineering and Biomedical Sciences, by the Litwin Foundation for
Alzheimer's Research, and by the NIH/NINDS award R01NS088040. 
Photo credit to Tom Deerinck and Mark Ellisman (National Center for Microscopy and Imaging Research) for \reffig{fig:model}(a).

\end{document}